\documentclass[12pt]{article}
\usepackage{amssymb}
\usepackage{graphicx}
\usepackage{amsmath}

\def\ee{\end{eqnarray}}

\def\underline{\underline}

\def\=:{=\hspace{-.7em}\raisebox{1.1ex}{.}\hspace{.1em}\raisebox{-0.2ex}{.} }

\newcommand {\beq}{\begin{eqnarray}}
\newcommand {\eeq}{\end{eqnarray}}
\newcommand {\non}{\nonumber\\}

\newcommand {\1}[1]{\frac{1}{#1}}

\newcommand {\eps}{\varepsilon}

\newcommand {\lam}{\lambda}

\newcommand{\vs}[1]{\vspace{#1 mm}}

\begin{document}
\begin{titlepage}
\begin{center}

\vs{10}
{\LARGE Baryonic Bound State of Vortices\\ 
in Multicomponent Superconductors}

\vs{10}
\vspace{0.5cm}
Muneto Nitta${}^{1}$,
Minoru Eto${}^{2}$, 
Toshiaki Fujimori${}^{3,4}$, 
Keisuke Ohashi${}^{5}$

\vspace{0.5cm}
{\it\small
${}^1$ Department of Physics, and Research and Education 
Center for Natural Sciences,}\\  
{\it\small 
 Keio University, Hiyoshi 4-1-1, Yokohama, Kanagawa 223-8521, Japan}\\
{\it\small 
${}^2$ Mathematical Physics Laboratory, Nishina Center, 
RIKEN, Saitama 351-0198, Japan}\\
{\it\small 
${}^3$ INFN, Sezione di Pisa, 
Largo B.~Pontecorvo, 3, 56127 Pisa, Italy}\\
{\it\small 
${}^4$ Department of Physics,``E. Fermi`'', University of Pisa, 
Largo B.~Pontecorvo, 3, 56127 Pisa, Italy}\\
${}^5$ {\it\small 
Department of Physics, Kyoto University, Kyoto 
606-8502, Japan}

\vspace{1cm}
{\bf Abstract}

\vspace{0.5cm}
\parbox{15cm}{
\small\hspace{15pt}
We construct a bound state of three 1/3-quantized Josephson coupled 
vortices in three-component superconductors with 
intrinsic Josephson couplings, 
which may be relevant with regard to iron-based superconductors.
We find a Y-shaped junction of three domain walls 
connecting the three vortices, resembling  
the baryonic bound state of three quarks in QCD. 
The appearance of the Y-junction (but not a $\Delta$-junction) 
implies that in both cases of superconductors and QCD, 
the bound state is described 
by a genuine three-body interaction 
(but not by the sum of two-body interactions).
We also discuss a confinement/deconfinement phase transition.
}
\end{center}

Keywords: multi-band superconductor, vortices, fractional flux quanta, Ginzburg-Landau free energy, interband phase difference soliton, 
confinement, baryon, meson, QCD.

\end{titlepage}

\setcounter{footnote}{0}
\renewcommand{\thefootnote}{\arabic{footnote}}

\section{Introduction}
Color confinement in quantum chromodynamics (QCD) 
is one of the most challenging 
problems in modern physics.
Quarks having fractional electric charges
$\pm(1/3)e$ or $\pm(2/3)e$ should be confined 
by color electric fluxes to form hadrons (mesons or baryons) 
with integer electric charges.
As a result, only mesons and baryons, made of 
two and three quarks, respectively, can be observed. 
A color electric flux tube  
stretched between a quark and an anti-quark
provides constant attractive force 
or a potential that is linearly dependent on 
the distance between the quarks. 
A recent lattice QCD simulation has confirmed 
this picture \cite{Bali:2000gf}.
On the other hand, one can imagine two possible 
configurations of color fluxes in a baryon: 
a Y-shaped junction or a $\Delta$-shaped junction 
connecting three quarks.
The $\Delta$-junction implies that the three-quark interaction 
is described by the sum of two-body interactions while 
the Y-junction implies a genuine three-body interaction. 
Although 
it has been a long standing issue as to which of 
these possible configurations is the actual one, 
a further study of lattice QCD simulations has clearly 
demonstrated the Y-junction \cite{Takahashi:2000te}.

A bound state of Josephson coupled vortices confined by domain walls  
exists in multicomponent superconductors, 
and it resembles the bound states of quarks 
confined by fluxes. 
Half-quantized vortices stably exist in 
two-band superconductors   
when the interband Josephson coupling is negligible \cite{Babaev:2002}.
When the interband Josephson coupling is taken into account, 
a half-quantized vortex is attached by a domain wall 
\cite{Tanaka:2001,Gurevich:2003} which 
extends to the edge of the sample. 
Since the domain wall pulls the vortex in order 
to reduce the energy, the half-quantized vortex is unstable. 
On the other hand, two half-quantized vortices 
winding around two different gap functions
are connected by a domain wall (sine-Gordon kink) 
\cite{Tanaka:2001,Gurevich:2003}
when the interband Josephson coupling is considered. 
The domain wall provides an attractive potential linearly depending 
on the distance between the two vortices \cite{Goryo:2007}; 
this bound state resembles a meson in QCD. 
In fact, confinement/deconfinement phase transition 
occurs at finite temperature, similar to the case in QCD, 
in which fluctuations of the domain wall contribute 
to entropy \cite{Goryo:2007}.
That is, vortices with fractional quanta 
are confined to become vortices with 
integer quanta below a certain temperature, 
while they are deconfined 
above that temperature.
However, the question is: 
Is there any model or actual material that indicates 
a baryonic bound state of vortices? 
In the absence of intrinsic Josephson terms, 
vortex bound states have been discussed in multi-component 
superconductors \cite{Smiseth:2004na}. 
How are they connected by domain walls, 
once the Josephson terms are turned on?

In this paper, we show that baryonic bound states 
indeed exist in three-component superconductors 
with intrinsic Josephson terms.
We show that three 1/3-quantized vortices are 
connected by a Y-shaped junction of domain walls, 
resembling a baryon in QCD.
We also discuss a confinement/deconfinement phase transition.
Above the certain critical temperatre, 
the three fractional vortices are 
deconfined from a vortex baryon.

Some two-band superconductors
exhibit type 1.5 superconductivity, i.e., 
repulsion at a short distance and 
attraction at a large distance 
between two integer vortices \cite{Babaev:2004hk}. 
Such a structure leads to a cluster of vortices, 
which has been experimentally confirmed 
in MgB$_2$ \cite{Moshchalkov:2009,Nishio:2010}.
Recently, three-component superconductors have attracted 
considerable attention 
because of the discovery of iron-based superconductors 
\cite{Hosono}. 
Therefore, we propose the multi-band superconductors 
to test baryonic bound states of vortices with 1/3 quanta 
which exhibit a confinement/deconfinement transition.
Our solution will also suggest the possibility of 
experimentally determining 
intrinsic Josephson couplings 
by determining the shape (angles and lengths) 
of a Y-junction of three vortices.

\section{Three-component superconductors}
Multicomponent ($n$-component) superconductors 
can be generically described by
the Ginzburg-Landau free energy, 
\beq
 F = \sum_{i=1}^n
\left[ {\hbar^2 \over 4m} 
\left|\left( \nabla + i{ 2e \over \hbar  c}{\bf A}\right)\Psi_i\right|^2
+ {\lam_i \over 4}(|\Psi_i|^2 -v_i^2)^2 \right]
+ F_{\rm J} + {{\bf H}^2 \over 8 \pi},
 \label{eq:free-energy}
\eeq
with the intrinsic Josephson terms 
\beq
 F_{\rm J} = 
 - \sum_{i\neq j} \1{2} \gamma_{ij}( \Psi_i^* \Psi_j +  \Psi_j^* \Psi_i)
= - \sum_{i\neq j}  \gamma_{ij} |\Psi_i||\Psi_j| 
  \cos (\theta_i - \theta_j).
  \label{eq:Josephson}
\eeq
Here, $\gamma_{ij}$ are constants and 
$\Psi_i$ is decomposed into amplitude and phase as 
$\Psi_i = |\Psi_i| e^{i \theta_i}$.
The phases of $\Psi_i$ and $\Psi_j$ preferably coincide 
for $\gamma_{ij}>0$
while they tend to have $\pi$ phase difference 
for $\gamma_{ij}<0$. 
All phases are the same in the ground state 
when $\gamma_{ij}>0$ for all $i$ and $j$, 
while the system is frustrated when $\gamma_{ij}<0$ 
for all $i$ and $j$.
The Hamiltonian (\ref{eq:free-energy}) 
is invariant under gauge symmetry, 
\beq
 {\bf A} \to {\bf A} - {\hbar c \over 2 e} \nabla\theta({\bf x}), \quad
 \Psi_i \to e^{i \theta({\bf x})} \Psi_i .
 \label{eq:gauge-tr}
\eeq
In the limit of $\gamma_{ij}=0$, 
the Hamiltonian (\ref{eq:free-energy}) enjoys 
$U(1)^n$ symmetry, 
of which the overall phase rotation exhibits 
the gauge symmetry (\ref{eq:gauge-tr}) 
while others exhibit global symmetry.
In this case, there appear $n-1$ Nambu-Goldstone modes. 
They are gapped for non-zero $\gamma_{ij}$ 
(the Legget modes).

\begin{figure}[h] 
\vs{-5}
\begin{center}
\includegraphics[width=5cm,clip]
{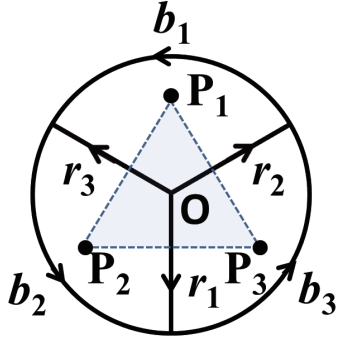}
\caption{
(Color online) 
The $(1,0,0),(0,1,0)$, and $(0,0,1)$ vortices 
are placed at P$_1$, P$_2$ and P$_3$, respectively. 
$b_i$ ($i=1,2,3$) corresponds to 1/3 circles at the boundary,   
and $r_i$ corresponds to the radial paths from the origin O 
to the circle at the boundary. 
The $(1,0,0)$, $(0,1,0)$, and $(0,0,1)$ vortices are 
encircled by $b_1 -r_3 + r_2$,  $b_2 -r_1 + r_3$, and 
$b_3 -r_2 + r_1$, respectively.
\label{fig:molecule}}
\end{center}
\end{figure}

Hereafter, we restrict ourselves to 
three-component ($n=3$) superconductors 
in two space dimensions \cite{Stanev:2010,Tanaka:2010}. 
We consider the positive Josephson couplings, 
$\gamma_{ij}>0$, while the negagive Josephson couplings, 
$\gamma_{ij}<0$, give frustrated systems \cite{Tanaka:2010}.
There exist three types of vortices, 
labeled as $(1,0,0),(0,1,0)$, and $(0,0,1)$,  
winding around the first,  second, and third component by $2\pi$, 
respectively. 
The energy of each vortex is logarithmically divergent 
when $\gamma_{ij}=0$ and linearly divergent when $\gamma_{ij}\neq 0$, 
if the system size is infinite.
Let us consider that the $(1,0,0),(0,1,0)$, and $(0,0,1)$ vortices 
are placed at the edges 
(P$_1$, P$_2$, and P$_3$, respectively) of 
a triangle, as shown in Fig.~\ref{fig:molecule}.
In the large circle, the total configuration is 
the integer vortex $(1,1,1)$, which implies 
that the total energy is finite.
In other words, the integer vortex has 
an internal structure made of 
three vortices, all of which are 1/3 quantized, as shown below.

Instead of the $U(1)^3$ generators $(1,0,0),(0,1,0)$, and $(0,0,1)$, 
let us prepare four linearly dependent generators: 
the gauge rotation $(1,1,1)$ and 
three gauge-invariant rotations $(0,-1,1)$, $(1,0,-1)$, and $(-1,1,0)$. 
Among these, only the gauge rotation is accompanied by 
gauge transformation (\ref{eq:gauge-tr}), 
while the others are all global phase rotations.
In these new generators, the winding of the $(1,0,0),(0,1,0)$, 
and $(0,0,1)$ vortices can be decomposed into
\beq
&&{\rm P}_1:\quad 
 (1,0,0)=\1{3}(1,1,1)    +0(0,-1,1)+\1{3}(1,0,-1)-\1{3}(-1,1,0),\non
&&{\rm P}_2:\quad 
 (0,1,0)=\1{3}(1,1,1)-\1{3}(0,-1,1)    +0(1,0,-1)+\1{3}(-1,1,0),\non
&&{\rm P}_3:\quad 
 (0,0,1)=\1{3}(1,1,1)+\1{3}(0,-1,1)-\1{3}(1,0,-1)+    0(-1,1,0).
\eeq
We see that all the paths $b_i$ ($i=1,2,3$) 
in Fig.~\ref{fig:molecule} correspond to $2\pi/3$ rotation of 
the gauge generator $(1,1,1)$ with Eq.~(\ref{eq:gauge-tr}) 
and consequently that 
these vortices are all 1/3 quantized; 
their magnetic flux is $\Phi_0/3$ with 
the unit flux quanta $\Phi_0=hc/2e$ \cite{footnote1}.
The $(1,0,0)$, $(0,1,0)$, and $(0,0,1)$ vortices are 
encircled by $b_1 -r_3 + r_2$,  $b_2 -r_1 + r_3$, and 
$b_3 -r_2 + r_1$, respectively (Fig.~\ref{fig:molecule}). 
Therefore, we can identify the paths 
$\pm r_1$, $\pm r_2$ and $\pm r_3$ 
corresponding to $\pm 2\pi/3$ of  
the global phase rotations $(0,-1,1),(1,0,-1)$, and $(-1,1,0)$, 
respectively.
The global phase rotation along the radial path $r_i$ 
can be written up to constant phases as 
\beq
&& r_1: \quad 
  \Psi_1 = |\Psi_1|,\quad 
  \Psi_2 = e^{-(2\pi i/3) f(r)}|\Psi_2|, \quad 
  \Psi_3 = e^{(2\pi i/3) f(r)}|\Psi_3| ,\non
&& r_2: \quad 
  \Psi_1 = e^{(2\pi i/3) f(r)}|\Psi_1| ,\quad 
  \Psi_2 = |\Psi_2|,\quad 
  \Psi_3 = e^{-(2\pi i/3) f(r)}|\Psi_3| ,\non
&& r_3: \quad 
  \Psi_1 = e^{-(2\pi i/3) f(r)}|\Psi_1| ,\quad 
  \Psi_2 = e^{(2\pi i/3) f(r)}|\Psi_2| ,\quad 
  \Psi_3 = |\Psi_3| ,\label{eq:r-pathes}
\eeq
where a function $f(r)$ has the boundary conditions 
$f(r=0)=1$ and $f(r\to \infty)=0$.

The integer vortex configuration (1,1,1) is 
of the Abrikosv type which has finite energy 
(due to the fact that the associated U(1) symmetry is
fully local), whereas the fractional vortices correspond to global
ungauged symmetries and hence they have a logarithmically divergent
energy, even in the absence of the Josephson terms, $\gamma_{ij}=0$. 
This is why the fractional vortices are confined 
whereas the integer vortices are
acceptable finite-energy solutions.

In the presence of the Josephson terms, $\gamma_{ij}\neq 0$,
we expect there to be a sine-Gordon kink in each path $r_i$ 
that connects two vortices. 
However, the question that arises is: 
how does it connect two vortices? 
Does it connect along the segment P$_j$P$_k$?

\section{Baryonic bound state}

\begin{figure}
\begin{center}
\includegraphics[width=10cm]{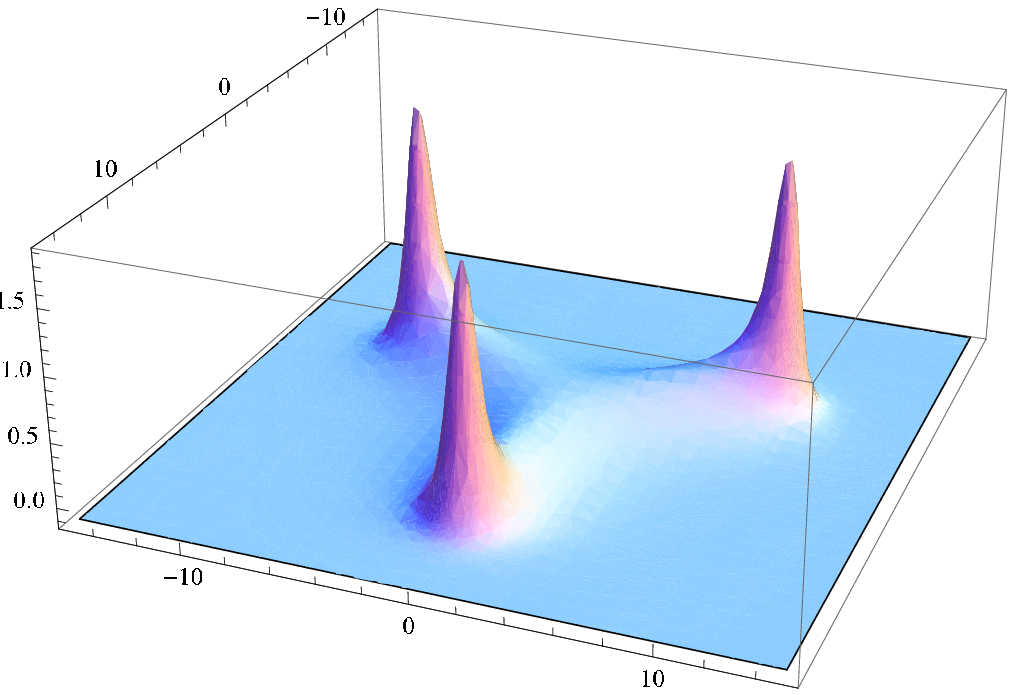}\\
(a)\\
\begin{tabular}{cc}
\includegraphics[width=5cm]{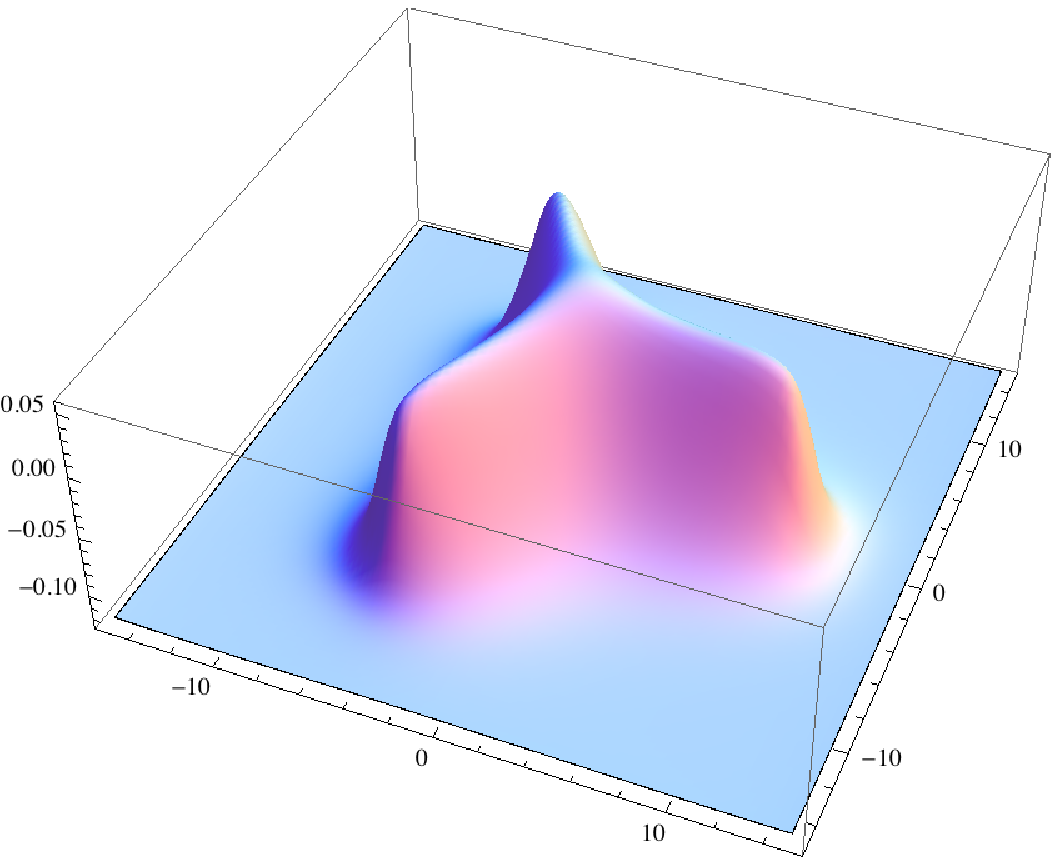} &
\includegraphics[width=5cm]{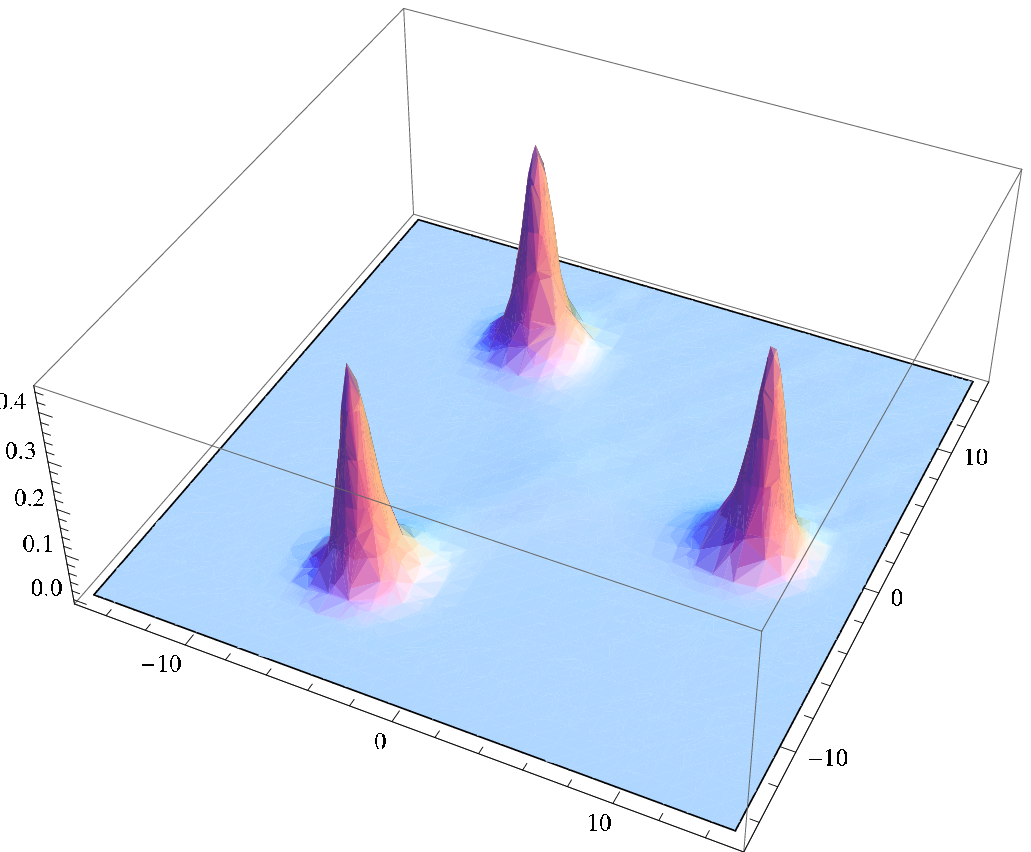} \\
(b) & (c)
\end{tabular}
\caption{(Color online) Baryonic bound state of vortices. 
Plots of (a) the total energy density, 
(b) the energy density of the Josephson couplings, 
and (c) the magnetic field. 
For simplicity, we take $\gamma_{ij}=\gamma>0$, 
but the general case is straightforward.
The relaxation method has been used with parameters 
$\hbar=c=2m=2e=v=\lam/2=1$ and $\gamma=0.02$.
\label{fig:baryon}}
\end{center}
\begin{center}
\begin{tabular}{cc}
\includegraphics[width=12cm]{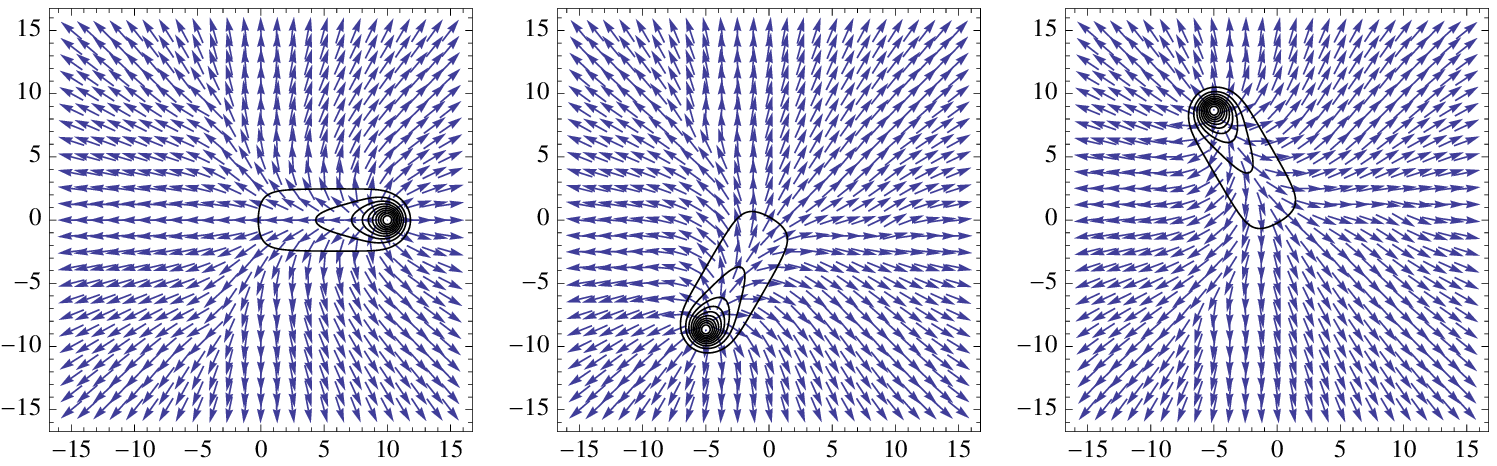}&
\includegraphics[width=5cm]{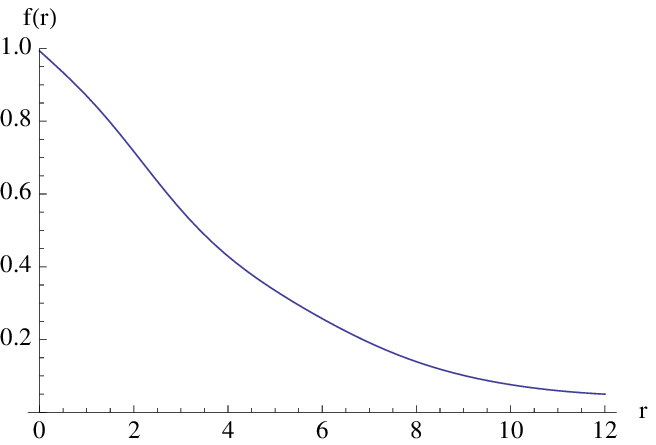}\\
(a) & (b)
\end{tabular}
\caption{
(Color online)
(a) The arrows and contour lines 
indicate, respectively,  the phases and amplitudes of 
$\Psi_1$(left), $\Psi_2$(middle) and $\Psi_3$(right).
(b) The plot of the function $f(r)$. 
\label{fig:baryon2}}
\end{center}
\end{figure}

We concentrate on the case of $\gamma_{ij}>0$ for all $i$ and $j$,  
which may be the case of iron-based superconductors. 
In this case, all phases are the same in the ground state, 
which is unique with respect to gauge transformation (\ref{eq:gauge-tr}). 
The phases at $r\to \infty$ 
are the same, which we consider to be zero according to the gauge symmetry; 
hence, Eq.~(\ref{eq:r-pathes}) holds, 
including for the constant phases. 
Along each radial path, the Josephson term in Eq.~(\ref{eq:Josephson}) 
can be written as
$\gamma_{ij} |\Psi_i| |\Psi_j| \cos((4\pi/3)f(r))$, 
from Eq.~(\ref{eq:r-pathes}). 
We thus find that it 
takes a non-zero value $\1{2} \gamma_{ij} |\Psi_i| |\Psi_j|$ 
at the center ($r=0$).
Therefore, the linear connection of vortices along the segments 
P$_j$P$_k$ due to the sine-Gordon kinks would increase energy  
because a domain (membrane) with finite energy 
appears inside the triangle P$_1$P$_2$P$_3$.
The sine-Gordon kinks should bend to form the Y-junction. 

In fact, the numerical solution indicates the Y-junction, 
as seen in Fig.~\ref{fig:baryon}. 
For simplicity, we have taken $\gamma_{ij}=\gamma \;(>0)$  but 
the general case is straightforward. 
In Fig.~\ref{fig:baryon}-(c), we find that the magnetic field 
is localized at the center of each vortex. 
Fig.~\ref{fig:baryon2}-(a) shows 
the phases of $\Psi_1$, $\Psi_2$, and $\Psi_3$, 
indicating a phase winding at P$_i$.
We also obtain the numerical solution for the function $f(r)$ 
in Eq.~(\ref{eq:r-pathes}), (Fig.~\ref{fig:baryon2}-(b)).
In this numerical simulation, we have fixed the positions 
of the three vortices. 
The wall tension leads to a linear potential (confining force), 
and these vortices collapse to form a single integer vortex.

\section{Confinement/deconfinement phase transition}
Here we discuss that the Y-junction can be stable 
at finite temperature 
and exhibits a phase transition, 
as proposed by Goryo {\it et.al.} \cite{Goryo:2007}
for a mesonic bound state of vortices in two-gap superconductors.
To this end, we construct the effective theory for the phases 
of the gap functions with keeping the amplitudes constants, 
$\Psi_i({\bf x})=v_i\exp(i \theta_i({\bf x}))$, given by
\beq
 F_{\rm eff.} =  \left[
\sum_i{\hbar^2v_i^2 \over 4m} 
\left(\nabla \theta_i + i{2e\over \hbar c} {\bf A}\right)^2 
+ \sum_{i\neq j} v_i v_j \gamma_{ij} (1-\cos (\theta_i - \theta_j))
\right], \label{eq:eff_model0}
\eeq
up to a constant. 
Let us define the phase diferences 
as gauge invariant dynamical variables as  
\beq
\phi_1 \equiv \theta_2 - \theta_3, \quad 
\phi_2 \equiv \theta_3 - \theta_1, \quad 
\phi_3 \equiv \theta_1 - \theta_2, \quad
 (\phi_1+\phi_2+\phi_3=0).
\eeq
With taking a  gauge fixing as 
\beq
 \theta_1 + \theta_2 + \theta_3 = 0 ,
\eeq  
the energy (\ref{eq:eff_model0}) is further reduced to 
\beq
 F_{\rm eff.} = \sum_i \left[
{\hbar^2v_i^2 \over 12 m} (\nabla \phi_i)^2 
+ \eta_i^2 (1-\cos \phi_i)
\right], \label{eq:eff_model}
\eeq
where we have set ${\bf A}=0$ and defined 
\beq
\eta_1^2 \equiv v_2 v_3 \gamma_{23},\quad 
\eta_2^2 \equiv v_3 v_1 \gamma_{31},\quad
\eta_3^2 \equiv v_1 v_2 \gamma_{12}.
\eeq

\begin{figure}[h]
\begin{center}
\begin{tabular}{cc}
\includegraphics[width=2cm]{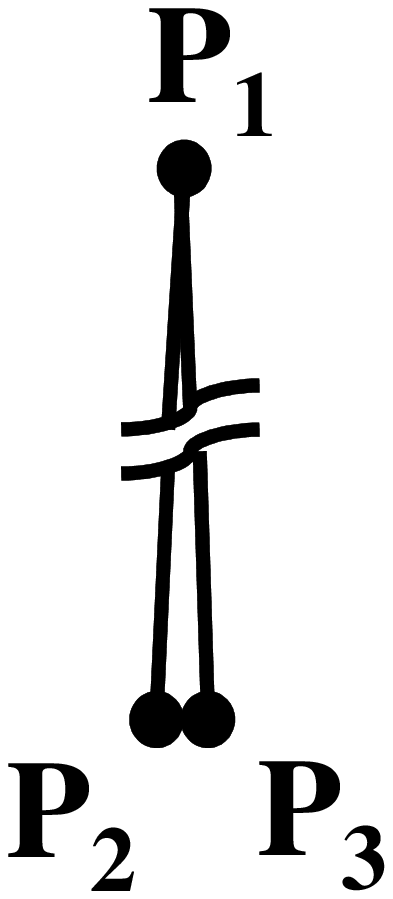} &
\includegraphics[width=5cm]{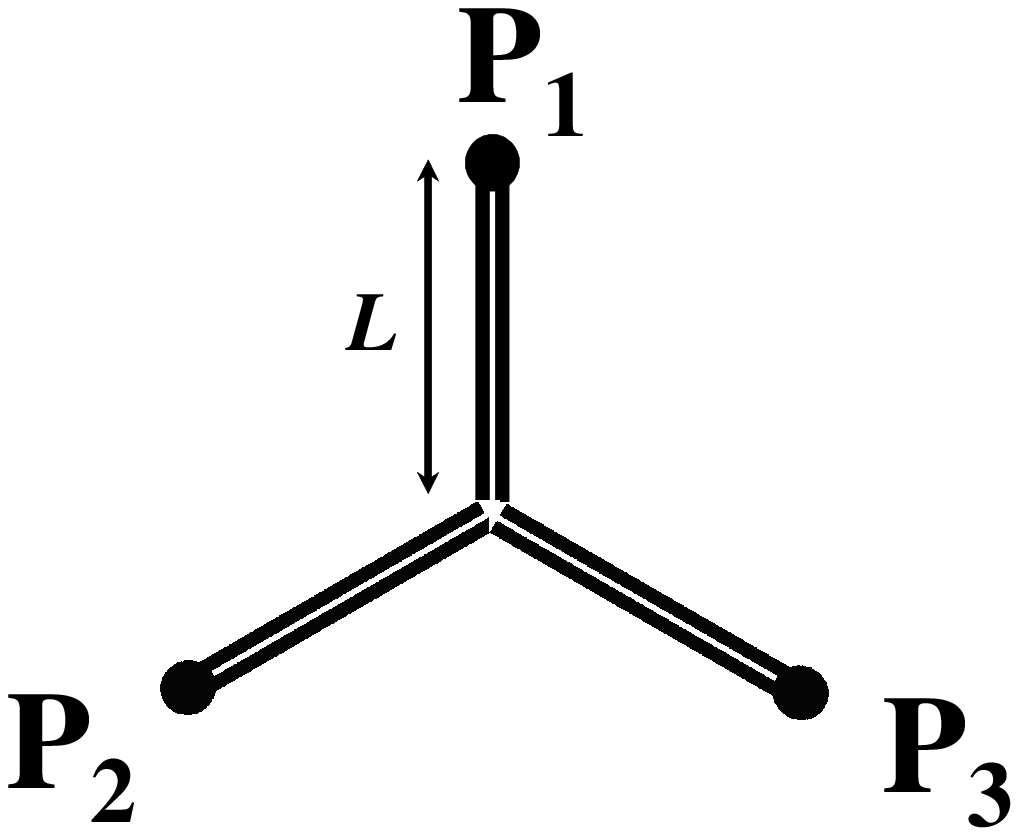}
\\
(a) & (b)
\end{tabular}
\caption{
(a) The two vortices P$_2$ $(0,1,0)$ 
and P$_3$ $(0,0,1)$ together are placed 
at the same position very far from 
the vortex P$_1$ $(1,0,0)$. 
They are connected by a sine-Gordon domain wall. 
(b) The most symmetric configuration at a finite temperature.
\label{fig:sine-Gordon}}
\end{center}
\end{figure}
In order to calculate the tension of the 
domain wall attached to 
the vortex $(1,0,0)$, 
let us place the two vortices P$_2$ $(0,1,0)$ 
and P$_3$ $(0,0,1)$ together
at the same position which is very far from 
the vortex P$_1$ $(1,0,0)$ as 
in Fig.~\ref{fig:sine-Gordon}(a).
In this situation, we can set $\theta_2=\theta_3$ so that we have 
\beq 
 \phi_1=0 ,\quad  \phi_2 = -\phi_3\equiv \phi.
\eeq
Then, the effective model (\ref{eq:eff_model}) reduces to 
the sine-Gordon model 
\beq
 F_{\rm eff.}^{(1)} = K^{(1)} (\nabla \phi)^2 
+ \frac{\Gamma^{(1)}}{2}(1-\cos \phi), \non
K^{(1)} \equiv {\hbar^2(v_2^2 + v_3^2) \over 12 m}
 , \quad 
\Gamma^{(1)} \equiv 2(\eta_2^2 + \eta_3^2 ).
\eeq
This can be rewritten as the Bogomol'nyi form 
\beq
 F_{\rm eff.}^{(1)} 
&=&  \left[K^{(1)} (\nabla \phi)^2 
         +  \Gamma^{(1)} \sin^2 ({\phi /2 })\right] \non
&=&   \left(\sqrt{K^{(1)}}\nabla \phi 
     \pm \sqrt{\Gamma^{(1)}}\sin ({\phi /2 })\right)^2
\mp 2 \sqrt{K^{(1)}\Gamma^{(1)}} \nabla \phi\sin ({\phi /2 })\non
&\geq&  \eps \label{eq:BPS}
\eeq
with the topological charge density 
in the second term in the second line,  
\beq
\eps \equiv  \mp 2 \sqrt{K^{(1)}\Gamma^{(1)}} \nabla \phi\sin ({\phi /2 })
  =   \pm 4 \sqrt{K^{(1)}\Gamma^{(1)}} \nabla \cos ({\phi /2 }).
\eeq
The most stable configurations with 
the minimum energy can be achieved by  
satisfying the Bogomol'nyi equation, 
obtained by $(....)^2 =0$ in the second line of 
Eq.~(\ref{eq:BPS}), {\it i.e.}
\beq
\sqrt{K^{(1)}}\nabla \phi 
\pm \sqrt{\Gamma^{(1)}}\sin ({\phi /2 }) =0 .
\eeq
One (anti-)kink solution can be obtained as
\beq
 \phi = 4 \arctan \exp  \left[\pm {1\over 4}\sqrt{\Gamma^{(1)} \over K^{(1)}} (x - x_0 )\right]  ,
\eeq
where $x$ is the coordinate perpendicular to the kink 
and $x_0$ denotes the position of the kink. 
The tension of the one (anti-)kink is
\beq 
 T^{(1)} 
&=& \left|\int d x \;\eps \right|
 = 4 \sqrt{K^{(1)}\Gamma^{(1)}} 
  \left| \left[\cos(\phi /2 )\right]_{x = -\infty}^{x= +\infty}\right| 
  = 8 \sqrt {K^{(1)}\Gamma^{(1)}} \non
&=& 8 \sqrt{{ \hbar^2(v_2^2 + v_3^2)(\eta_2^2 + \eta_3^2)}
               \over 12 m}
= 8 \hbar \sqrt{{v_1(v_2^2 + v_3^2) 
        (v_2 \gamma_{21} + v_3 \gamma_{31}) 
        \over 6 m}}.
\eeq

In  the same way, the other two domain walls attached 
to the $(0,1,0)$ and $(0,0,1)$ have the tensions
\beq
&& T^{(2)} = 8 \sqrt {K^{(2)}\Gamma^{(2)}} 
, \quad 
 K^{(2)} \equiv {\hbar^2(v_3^2 + v_1^2) \over 12m}, \quad 
 \Gamma^{(2)} \equiv 2(\eta_3^2 + \eta_1^2), \non
&& T^{(3)} = 8 \sqrt {K^{(3)}\Gamma^{(3)}}
, \quad 
 K^{(3)} \equiv {\hbar^2(v_1^2 + v_2^2) \over 12m}, \quad 
 \Gamma^{(3)} \equiv 2(\eta_1^2 + \eta_2^2) ,
\eeq
respectively. 

We are now ready to discuss the confinement/deconfinement 
phase transition.
For simplicity, we consider the most symmetric case with 
\beq 
v_1=v_2=v_3\equiv v ,\quad
 \gamma_{12}=\gamma_{23}=\gamma_{31}\equiv \gamma. 
\label{eq:symmetric}
\eeq
The tension of each domain wall becomes 
\beq
 T_{\rm dw} = 8 \hbar v^2 \sqrt{{2\gamma \over 3m}}.
\eeq
In this case, the molecule is ${\bf Z}_3$ symmetric 
as in Fig.~\ref{fig:sine-Gordon}(b). 
For each domain wall with the length $L$, 
the total energy and entropy can be evaluated as 
\cite{Goryo:2007,Kogut:1979wt}
\beq
 E = T_{\rm dw}L, \quad
 S_{\rm dw} = {k_B \ln 2\pi \over \xi}L,
\eeq
respectively, with a short lengh cut-off 
$\xi$ which is the largest among 
the coherence length and the penetration depth.  
Consequently, 
the free energy of each domain wall 
at the temperature $T$ is given by 
\beq
 F_{\rm dw} &=& E_{\rm dw} -TS_{\rm dw}  
= \left(T_{\rm dw}- {k_B T \ln 2\pi \over \xi}\right) L 
= AL, \\
A &\equiv& T_{\rm dw}- {k_B T \ln 2\pi \over \xi}.
\eeq
When the coefficient $A$ is positive, the interger vortex 
is stable, {\it i.e.}, in the confinement phase. 
On the other hand, 
when the coefficient $A$ is negative, the integer vortex 
tends to be split into a set of the three fractional vortices 
in order to reduce the free energy of the domain walls, 
that is, the deconfinement occurs. 
Therefore, the critical temperature for the deconfinement is 
found to be 
\beq
 T_{\rm crit} = {\xi T_{\rm dw}\over  k_B \ln 2\pi} 
= {16 \xi \hbar v^2 \over k_B \ln 2\pi}\sqrt{{2\gamma \over 3m}}.
\eeq
This expression is the same with Goryo {\it et.al}~\cite{Goryo:2007}.
In the most symmetric case 
with Eq.~(\ref{eq:symmetric}) which we are considering, 
the confinement mechanism is essentially the same with 
the case of two-gap superconductors.

\section{Summary and Discussion}
In summary, we have constructed baryonic states 
of three 1/3-quantized vortices and have 
found that these vortices are connected by the Y-junction 
of domain walls, 
resembling a baryonic bound state of three quarks in QCD.
In both cases of superconductors and QCD, 
the appearance of the Y-junction and not of a $\Delta$-junction 
implies that the bound state is described by a genuine 
three-body interaction and not by the sum of two-body 
interactions. 
The confinement/deconfinement transition of vortices 
has been studied. 
This common feature between superconductors and QCD 
will shed a new light on 
the color confinement problem of QCD. 

The similarities between superconductors and QCD 
should be further clarified. 
As a toy model of QCD, 
Shifman and Unsal \cite{Shifman:2008yb} 
considered an $SU(2)$ gauge theory  
in three space dimensions 
with one direction compactified  
as ${\bf R}^2 \times S^1$.
The theory becomes a $U(1)$ gauge theory 
in two dimensional space ${\bf R}^2$ 
in the limit of a small radius of $S^1$.
It was shown by Polyakov \cite{Polyakov:1976fu} 
that the confinement occurs in a $U(1)$ gauge theory in 
two dimensional space; 
electrically charged particles (quarks) 
are confined by an electric flux. 
By taking a duality, quarks are mapped into vortices, 
while electric fluxes are mapped into sine-Gordon 
domain walls, so that 
a meson made of two quarks is mapped 
to a mesonic bound state of two vortices. 
There, the quark confinement can be understood 
as the vortex confinement which 
is described by two-gap superconductors. 
The confinement is nontrivial in the former, 
while it can be easily shown in the latter. 
We therefore expect that a discussion along the same line 
shows that a baryon made of three quarks 
in $SU(3)$ QCD is mapped to 
a bayonic bound state of 
three vortices found in this paper. 
We conjecture that  
the existence of a $Y$-junction in a baryon in QCD 
can be shown by using a duality map to 
three-gap superconductors.
In multi-gap superconductors with 
more than three gaps, the bound states of more vortices 
should exist which 
may correspond to tetraquarks, pentaquarks, etc. 
in QCD.  

Although we have chosen  
the same Josephson couplings $\gamma_{ij}=\gamma$ 
as an example, 
an extension to the general case is straightforward.
It may be useful  
to determine intrinsic Josephson couplings 
of multicomponent superconductors such as iron-based superconductors 
by determining the shape of the Y-junction. 
Several interesting studies have been conducted on 
for vortex mesons in two-component 
or $p$-wave superconductors, 
for instance, the studies on a lattice of vortex mesons 
and twistons \cite{Tanaka:2008} and 
those on vortex clusters \cite{Moshchalkov:2009,Nishio:2010}.
We hope that our work will stimulate further 
theoretical and experimental studies of 
multicomponent superconductors, 
particularly iron-based ones.

\section*{Acknowledgement}
M.~N. would like to thank D.~Inotani for useful comments.
The work of M.~N. is supported in part by 
Grant-in Aid for Scientific Research (No. 23740198) 
and by the ``Topological Quantum Phenomena'' 
Grant-in Aid for Scientific Research 
on Innovative Areas (No. 23103515)  
from the Ministry of Education, Culture, Sports, Science and Technology 
(MEXT) of Japan. 
The work of M.E. is supported in part by 
Grant-in Aid for Scientific Research (No. 23740226) and 
by Japan Society for the Promotion of Science (JSPS) 
and Academy of
Sciences of the Czech Republic (ASCR) under the Japan - Czech Republic Research Cooperative
Program


\newcommand{\J}[4]{{\sl #1} {\bf #2} (#3) #4}
\newcommand{\andJ}[3]{{\bf #1} (#2) #3}
\newcommand{\AP}{Ann.\ Phys.\ (N.Y.)}
\newcommand{\MPL}{Mod.\ Phys.\ Lett.}
\newcommand{\NP}{Nucl.\ Phys.}
\newcommand{\PL}{Phys.\ Lett.}
\newcommand{\PR}{ Phys.\ Rev.}
\newcommand{\PRL}{Phys.\ Rev.\ Lett.}
\newcommand{\PTP}{Prog.\ Theor.\ Phys.}
\newcommand{\hep}[1]{{\tt hep-th/{#1}}}


\end{document}